\documentclass[twocolumn]{aastex6}
\usepackage{graphicx}
\usepackage{amsmath}
\usepackage{natbib}

\usepackage{color}

\begin{document} 

\title{On the origin of dynamically isolated hot Earths}

\author{Arieh K\"onigl\altaffilmark{1}, Steven Giacalone\altaffilmark{2}, and Titos 
Matsakos\altaffilmark{1}}

\altaffiltext{1}{Department of Astronomy \& Astrophysics and The Enrico Fermi Institute,
The University of Chicago, Chicago, IL 60637, USA}
\altaffiltext{2}{Department of Physics, The University of Chicago, Chicago, IL 60637, USA}

\shortauthors{K\"onigl, Giacalone, \& Matsakos}
\shorttitle{Origin of isolated hot Earths}

\begin{abstract} 
A distinct population of planetary systems that contain dynamically isolated, Earth-size planets with orbital periods $P_\mathrm{orb}\sim1$\,day was recently identified in an analysis of data from the \textit{Kepler} planet candidate catalog. We argue that these objects could represent the remnant rocky cores of giant planets that arrived at the stellar vicinity on high-eccentricity orbits and were rapidly stripped of their gaseous envelopes after crossing their respective Roche limits (RLs) $a_\mathrm{R,p}$. In this picture, objects with $P_\mathrm{orb}\gtrsim1$\,day are mostly ``early'' cores that originated in planets with an initial periastron distance $a_\mathrm{per,0}\le a_\mathrm{R,p}$; they had high initial eccentricities but their orbits underwent fast tidal circularization after the cores were exposed. Objects with  $P_\mathrm{orb}\la 1$\,day are, by contrast, mostly ``late'' cores that originated in planets with $a_\mathrm{per,0}>a_\mathrm{R,p}$; these planets underwent orbital circularization to a radius $>a_\mathrm{per,0}$ but eventually reached $a_\mathrm{R,p}$ through tidal orbital decay. This picture naturally accounts for the spatial distribution of hot Earths and for the similarity of their inferred occurrence rate to that of hot Jupiters, and it fits well with the interpretation of the so-called sub-Jovian desert in the orbital-period--planetary-mass plane in terms of high-eccentricity planet migration to the vicinity of the RL.
\end{abstract}


\maketitle

\section{Introduction}
\label{sec:intro}

Our understanding of how extrasolar planets form and evolve has benefited from the accumulation of large data sets, which made it possible to identify distinct populations of exoplanets (e.g., hot Jupiters, HJs) and to conduct detailed studies of select groups of objects (such as the ultra-short-period, USP, planets---defined by having orbital periods $P_\mathrm{orb}<1$\,day---that were detected by the \textit{Kepler Space Telescope}; \citealt{Sanchis-Ojeda+14}). Recently, \citet[][hereafter SC16]{SteffenCoughlin16} analyzed data from the \textit{Kepler} planet candidate catalog and reported identifying a distinct population of Earth-size planets with $P_\mathrm{orb}\sim1$\,day, found predominantly among systems that contain either a single planet or an effectively single one (an inner planet that has no planetary companion with a period ratio $\la 6$ and is therefore dynamically isolated), which they dubbed ``hot Earths'' (HEs; the HE sample used by SC16 comprises planet candidates with $P_\mathrm{orb}<2$\,days and radii $R_\mathrm{p}<2\,R_\earth$,  where the subscript $\mathrm{p}$ denotes a planet). SC16 inferred that the occurrence rate of HEs is at least $\sim$1/3 of that of HJs and argued that the actual rate (after accounting for sample incompleteness) is roughly twice as large and could be comparable to that of HJs ($\sim$0.5\%  in the \textit{Kepler} database; e.g., \citealt{Santerne+16}). It is noteworthy that the estimated occurrence rate of the USP planets discussed by \citet{Sanchis-Ojeda+14} is also similar to that of HJs and that almost all of these planets also have radii $<2\,R_\earth$; however, as pointed out by SC16, the precise degree of overlap between the \textit{Kepler} USP planet set and the newly identified HE population remains to be determined. In any case, the HEs are not a subset of USP planets because the HE population---which, based on Figure~3 of SC16, roughly occupies the $\log{P_\mathrm{orb}(\mathrm{days})}$ interval $(-0.3,0.3)$ with a peak near $0$---extends beyond the nominal orbital range of the USP planet set ($P_\mathrm{orb}<1\,$day).

The fact that HJs are also dynamically isolated objects and the similarity between their occurrence rate and that of HEs make it natural to consider a possible link between these two types of close-in planets \citep[SC16; see also][]{SteffenFarr13}. One intriguing possibility, suggested by \citet{Valsecchi+14}, is that close-in rocky planets may represent the remnant cores of HJs that lost their gaseous envelopes through Roche lobe overflow (RLO) after their orbits decayed (due to tidal interaction with the star) down to the Roche limit (RL). However, after considering this scenario more closely, \citeauthor{Valsecchi+15} (\citeyear{Valsecchi+15}; see also \citealt{Jackson+16} and \citealt{GinzburgSari17}) concluded that it cannot account for the observed USP planets. They modeled the evolution of HJs that undergo RLO under the assumption that the mass transfer from the planet is stable on a dynamical timescale. In this case, the duration of the mass transfer is relatively long ($\gtrsim 1$\,Gyr), and the decrease in the planet's mass $M_\mathrm{p}$ leads initially to an increase in $P_\mathrm{orb}$. This outward motion is reversed when $M_\mathrm{p}$ drops to a value below which $R_\mathrm{p}$ decreases rapidly with decreasing $M_\mathrm{p}$, with the turnaround value being higher the larger the mass of the planet's core. This picture implies an inverse correlation between the mass of a remnant core and its orbital period, and \citet{Valsecchi+15} determined that the remnant masses predicted by this model for $P_\mathrm{orb}<1$\,day are inconsistent with the small sizes of the observed USP planets. A similar reasoning could be used to infer that this picture also cannot account for the observed HEs. An independent argument against such an interpretation is that the mass-loss timescale for dynamically stable RLO (which equals the timescale $\tau_\mathrm{d,p}$ for tidal decay of the orbit and thus scales inversely with $M_\mathrm{p}$; see Equation~\eqref{eq:a_t}) will become longer than the system's age well before the parent planet's mass drops to $\sim$1\,$M_\earth$ \citep[see][]{GinzburgSari17}.

In this Letter we revisit the ``remnant cores'' scenario in the context of the high-eccentricity migration (HEM) model for the origin of HJs and under the assumption that the RLO process in HJs is dynamically unstable. In the HEM picture \citep{FordRasio06}, a planet with a comparatively large initial value of $P_\mathrm{orb}$ is placed on a high-eccentricity orbit following a sudden planet--planet scattering event or through a slower interaction such as Kozai migration or secular chaos; this brings the planet close enough to the star for tidal interaction (predominantly involving dissipation in the planet) to result in orbital circularization at an orbital period within the HJ range ($P_\mathrm{orb}<7$\,days; the orbital period can be further reduced through tidal dissipation in the star on the timescale $\tau_\mathrm{d,p}$, which is typically significantly longer than the circularization time $\tau_\mathrm{cir,p}$). This picture is consistent with the observed $P_\mathrm{orb}$ distribution of HJs \citep[e.g.,][]{Matsumura+10,ValsecchiRasio14}. Furthermore, \citet[][hereafter MK16]{MatsakosKonigl16} demonstrated that this model leads to a natural interpretation of the dearth of sub-Jupiter-mass planets on short-period orbits (the so-called sub-Jovian desert) in the planetary period--mass plane, and \citet[][hereafter GMK17]{Giacalone+17} showed that the eccentricity gradient predicted in this picture near the orbital circularization locus in the $M_\mathrm{p}$--$P_\mathrm{orb}$ plane is likewise consistent with the data. The adoption of a rapid (dynamically unstable) mass transfer prescription in modeling RLO from a Jupiter-mass planet was recently advocated by \citet{JiaSpruit17} on the grounds that, in contrast with binary stars, most of the angular momentum removed in this case by the mass-transferring stream is probably not deposited in an accretion disk but is instead absorbed by the host star. Tidal interaction models that incorporate the assumption of unstable RLO can account for the orbital distribution of HJs \citep{Jackson+09} and for the dearth of close-in planets around fast-rotating stars \citep{TeitlerKonigl14}. This assumption is also central to the ``stranded HJ'' interpretation of the spin--orbit alignment properties of planets around cool and hot stars \citep{MatsakosKonigl15} and to the sub-Jovian desert model of MK16.

MK16 and GMK17 considered the tidal evolution of planets that arrive at the stellar vicinity through HEM but remain outside their respective RLs. Here, we focus attention on HJs that arrive in this manner but end up crossing the RL radius $a_\mathrm{R,p}$ (Equation~\eqref{eq:RLp}).\footnote{Although we only refer explicitly to HJs, our proposed scenario also encompasses lower-mass giant planets that undergo HEM.} This can happen in one of two ways: (1) the initial (subscript 0) periastron distance $a_\mathrm{per,0}=(1-e_0)\,a_0$ (where $e$ is the eccentricity and $a$ is the semimajor axis) is $\le a_\mathrm{R,p}$, or (2) the planet arrives with $a_\mathrm{per,0}>a_\mathrm{R,p}$ and its orbit is circularized to $a_\mathrm{cir}\approx(1+e_0)\,a_\mathrm{per,0}$ (estimated by assuming conservation of specific angular momentum; \citealt{FordRasio06}), but subsequent orbital decay reduces its semimajor axis to the RL value. In both cases, we assume that, after crossing the RL, the planet loses its gaseous envelope and is converted into a rocky remnant of size $R_\mathrm{c}$ and mass $M_\mathrm{c}$, which we identify with the planet's core (subscript $\mathrm{c}$). This conversion is postulated to occur rapidly---on a timescale that, in particular, is $\ll \tau_\mathrm{cir,c}$ in case~(1) and $\ll \tau_\mathrm{d,c}$ in case~(2). In the first case, this is the predicted effect of the strong tidal force that characterizes the region inside the RL. A potential complication in this case is that most of the tidal energy that is extracted near periastron is deposited in the planet and could in some instances lead to the unbinding of its orbit \citep[e.g.,][]{Faber+05}. However, for the parameter values that we consider, we expect the planet to remain bound even as it is progressively reduced in size over successive periastron passages \citep[see][]{Guillochon+11}. Although the remnant core in this case will initially possess the parent planet's high eccentricity, its orbit will be circularized on a timescale $\tau_\mathrm{cir,c}$ that is possibly even shorter than the circularization time $\tau_\mathrm{cir,p}$ for the parent planet (see Section~\ref{sec:model}). In the second case, the rapid emergence of a remnant core on a circular orbit of radius $a_\mathrm{R,p}$ is the expected outcome of a dynamically unstable RLO \citep[see][]{JiaSpruit17}. Since the cores in case~(1) are exposed within several periastron passages after their parent planets are placed on a high-$e_0$ orbit, whereas the cores in case~(2) emerge much later (on a timescale $\tau_\mathrm{d,p} \gtrsim 1$\,Gyr), we henceforth refer to them as ``early'' and ``late'' cores, respectively. In both cases, the cores will spiral inward after they become exposed, but, because of their lower mass and weaker expected tidal coupling to the star, their orbital decay rates will be much lower than those of their parent planets (see Equation~\eqref{eq:a_t}). As we demonstrate below, this picture provides a consistent explanation of the origin of isolated HEs.

\section{Modeling Approach}
\label{sec:model}

We follow the formulation presented in MK16 and updated in GMK17, and these references should be consulted for further details---including in particular the initial distribution of $a_0$ and the distributions of $R_\mathrm{p}$, $M_\mathrm{p}$, $t_\mathrm{age}$ (the system's age), $t_\mathrm{arr}$ (the planet's arrival time at the stellar vicinity), and $P_*$ (the stellar rotation period; the subscript\,$*$ denotes the host, which we take to be a Sun-like star). We carry out Monte Carlo simulations of planets that arrive on high-$e_0$ orbits and become either HJs or ``early'' cores depending on whether the ratio $a_\mathrm{per,0}/a_\mathrm{R,p}$ is $>$\,1 or $\le1$. The planet's RL is given by
\begin{equation}
  a_\mathrm{R,p}=q\,(M_*/M_\mathrm{p})^{1/3}\,R_\mathrm{p}\,,\label{eq:RLp}
\end{equation}
where $q$ is a numerical coefficient that we take to be either 3.46, the best-fit value obtained by MK16 from modeling the shape of the sub-Jovian desert boundary, or 2.7, the lower limit derived from the hydrodynamic simulations of \citealt{Guillochon+11}.\footnote{If the planet arrives through a diffusive process such as secular chaos, HEM may be stopped before the periastron distance drops below $a_\mathrm{R,p}$ \citep[see][]{WuLithwick11}. In this work, we assume that, even if this were to happen, other HEM processes (notably planet--planet scattering) would act to populate orbits with $a_\mathrm{per,0}\le a_\mathrm{R,p}$.} The values of $e_0$ are chosen from a distribution of the form $\partial{f}/\partial e_0\propto10^{-\beta e_0}$ that extends between $e_\mathrm{0,min}$ and $e_\mathrm{0,max}$; we adopt $e_\mathrm{0,min}=0.5$ and show results for $\beta=0$ and~2 (which correspond to two of the distributions plotted in figure~4 of \citealt{WuLithwick11}) and for $e_\mathrm{0,max}=0.9$ and~0.95 (where~0.95 is close to the critical value of $e_0$ above which, based on the results of \citealt{Guillochon+11}, a planet with $a_\mathrm{per,0}\le a_\mathrm{R,p}$ would be ejected). We sample the planetary mass and radius as independent quantities, with $M_\mathrm{p}$ in the nominal HJ mass range ($0.3$--$10\,M_\mathrm{J}$) and $R_\mathrm{p}$ in the corresponding planetary radius range (taken to be $9$--$20\,R_\earth$). In modeling the remnant cores, we sample radii from a uniform distribution in the range $0.8$--$1.25\,R_\earth$ (the Earth-type size range adopted in \citealt{Sanchis-Ojeda+14}) and assign mass values assuming a mean density $\rho_\mathrm{c}=6\,\mathrm{g\,cm}^{-3}$.\footnote{The mean density of a solid planet with a mantle-and-core structure similar to that of Earth is expected to increase with radius for the $R_\mathrm{c}$ values under consideration \citep[e.g.,][]{Seager+07}. In our constant-density approximation, we use a value of $\rho_\mathrm{c}$ that is near the lower end of the predicted range.} The RL for remnant cores is determined  from  the expression for the corresponding orbital period given in \citet{Sanchis-Ojeda+14},
\begin{equation}
  (P_\mathrm{orb})_\mathrm{R,c}=11.3\,\rho_\mathrm{c}^{-1/2}\,\mathrm{hr}\,.\label{eq:RLc}
\end{equation}

The semimajor axis and eccentricity of any sampled planet whose circularization radius $a_\mathrm{cir}$ corresponds to $P_\mathrm{orb}\la 7$\,days are tidally evolved under the assumption that energy dissipation in the planet rapidly leads to an alignment of its spin and orbital angular momenta and to a pseudosynchronization of its rotational and orbital periods (so that the rate of change of the rotation period is zero). We also assume, for simplicity, that the stellar spin is aligned with the orbit, and we neglect its variation with time. The evolution equations are then given by
\begin{equation}
\label{eq:a_t}
\begin{split}
  \frac{da}{dt}=9\frac{n}{Q^\prime_{\mathrm{p}}}\frac{M_*}{M_{\mathrm{p}}}\frac{R_{\mathrm{p}}^5}{a^4}(1-e^2)^{-15/2}\\ \times\left[\frac{(\textit{f}_2(e^2))^2}{\textit{f}_5(e^2)}-\textit{f}_1(e^2) \right]\\
+9\frac{n}{Q^\prime_\mathrm{*,p}}\frac{M_\mathrm{p}}{M_*}\frac{R_*^5}{a^4}(1-e^2)^{-15/2}\\ \times\left[\textit{f}_2(e^2)\frac{\omega_*}{n}(1-e^2)^{3/2}-\textit{f}_1(e^2) \right]\,,
\end{split}
\end{equation}
\begin{equation}
\label{eq:e_t}
\begin{split}
  \frac{de}{dt}=\frac{81}{2}\frac{n}{Q^\prime_{\mathrm{p}}}\frac{M_*}{M_{\mathrm{p}}}\frac{R_{\mathrm{p}}^5}{a^5}e(1-e^2)^{-13/2}\\ \times\left[\frac{11}{18}\frac{\textit{f}_4(e^2)\textit{f}_2(e^2)}{\textit{f}_5(e^2)}-\textit{f}_3(e^2)\right]\\
+\frac{81}{2}\frac{n}{Q^\prime_\mathrm{*,p}}\frac{M_\mathrm{p}}{M_*}\frac{R_*^5}{a^5}e(1-e^2)^{-13/2}\\ \times\left[\frac{11}{18}\textit{f}_4(e^2)\frac{\omega_*}{n}(1-e^2)^{3/2}-\textit{f}_3(e^2) \right]
\end{split}
\end{equation}
(e.g., \citealt{Matsumura+10}, where one can also find the expressions for the eccentricity functions\footnote{Note that the value of each of these functions is 1 at $e=0$.}  $f_1$,\,.\,.\,.,\,$f_5$). In Equation~\eqref{eq:a_t}, $n=(GM_*/a^3)^{1/2}$ (where $G$ is the gravitational constant) is the mean motion and $\omega_*\equiv2\pi/P_*$. Following GMK17 (see also \citealt{Matsumura+10}), we write the (modified) planetary tidal quality factor as $Q^\prime_\mathrm{p}=Q^\prime_{\mathrm{p}1}(P_\mathrm{orb}/P_1)$, with $Q^\prime_{\mathrm{p}1}=10^6$ and $P_1=4$\,days. We use the same form (and the same value of $P_1$) to approximate the stellar tidal quality factor for planet-induced dissipation, $Q^\prime_{*,\mathrm{p}}$, and set  $Q^\prime_{*,\mathrm{p}1}=10^6$. For any given planet (identified by the values of $a_0$, $e_0$, $R_\mathrm{p}$, and $M_\mathrm{p}$) we integrate Equations~\eqref{eq:a_t} and~\eqref{eq:e_t} over the time interval ($t_\mathrm{age}-t_\mathrm{arr}$).

Once a planet is converted into a rocky remnant, we replace the values of $M_\mathrm{p}$, $R_\mathrm{p}$, $Q^\prime_\mathrm{p}$, and $Q^\prime_{*,\mathrm{p}}$ by the corresponding core quantities. Using $Q^\prime_\mathrm{c}\sim100$ \citep[e.g.,][]{ClausenTilgner15}, we estimate that the circularization time $\tau_\mathrm{cir,c}\propto Q^\prime_\mathrm{c}\,M_\mathrm{c}/R_\mathrm{c}^5$ (see Equation~\eqref{eq:e_t}) of an ``early'' core is even shorter than the time it would have taken the orbit of its parent planet to be circularized. It is therefore a good approximation to assume that, after emerging at the parent planet's pericenter, such a core is transported ``instantaneously'' to the circularization radius $a_\mathrm{cir}(a_0,e_0)$ and that its post-formation evolution is similar to that of a ``late'' core in having $e\approx0$. Thus, for both early and late cores, only Equation~\eqref{eq:a_t}---in which just the now-dominant stellar dissipation term is retained---is employed to model the orbital evolution. Because of their significantly lower masses, remnant cores might induce a much weaker tidal dissipation than HJs \citep[e.g.,][]{EssickWeinberg16}. We account for this possibility by setting $Q^\prime_{*,\mathrm{c}}=10^7$, but we also examine the consequences of continuing to use our adopted expression for $Q^\prime_{*,\mathrm{p}}$ in this calculation. The evolution is followed until either the total allotted tidal interaction time ($t_\mathrm{age}-t_\mathrm{arr}$) is exceeded or the core reaches its RL (Equation~\eqref{eq:RLc}) and is removed from the system.\footnote{\citet{JiaSpruit17} argued that the RLO process may be dynamically stable for rocky planets with $M_\mathrm{c}<M_\earth$, which would prolong their dispersal times. However, any such modification would not significantly affect our results.} 

\section{Results}
\label{sec:results}

\begin{figure*}
  \includegraphics[width=\textwidth]{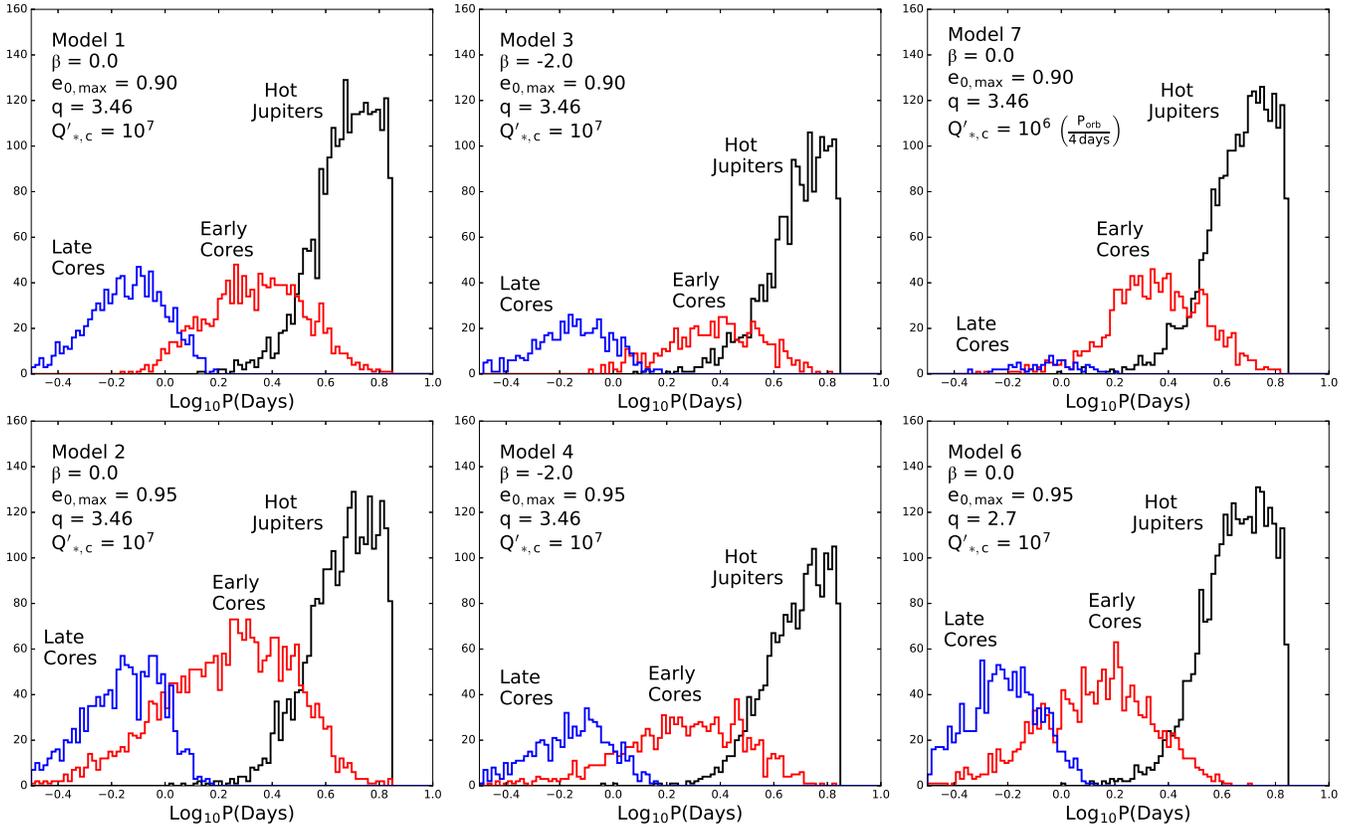}
  \caption{
$P_\mathrm{orb}$ distribution of the surviving HJs and remnant cores for several of the models  presented in Table~\ref{tab:table1}. The number and defining parameters of the relevant model are listed in each panel.
} \label{fig:fig1}
\end{figure*}

We conducted Monte Carlo experiments, each involving 30,000 planet drawings, for various combinations of the parameters $\beta$, $e_\mathrm{0,max}$, $q$, and $Q^\prime_{*,\mathrm{c}}$. The results for eight of these models are summarized in Table~\ref{tab:table1}. The number of surviving HJs (i.e., HJs that did not cross their respective RLs) as well as those of surviving and destroyed cores (both early and late) are listed as percentages of the total number of sampled planets that satisfy $t_\mathrm{age}\ge t_\mathrm{arr}$.\footnote{For each of the models presented in Table~\ref{tab:table1} we conducted 10 runs, which we verified was a sufficient number for obtaining the mean and standard deviation of the quantities listed in columns 2--6 to an accuracy of one decimal place. The models included in the table all have values in these columns that lie within one standard deviation of the respective means.}  These percentages do not have a direct astrophysical significance and are useful only insofar as they convey information on the relative abundances of the different types of objects that we model. The $P_\mathrm{orb}$ distributions of the surviving HJs and cores are shown in Figure~\ref{fig:fig1} for six of these cases.

Model~1 corresponds to the parameter combination that was used in GMK17 to demonstrate the compatibility of the HEM scenario with the eccentricity properties of close-in planets, and employs $Q^\prime_{*,\mathrm{c}}=10^7$. This model yields comparable numbers of surviving HJs and remnant cores, consistent with the inference of SC16 about the relative numbers of HEs and HJs.\footnote{SC16 defined HJs somewhat differently than we do, but the fractional difference in the number of objects selected from the \textit{Kepler} database using these two definitions is not large.} Furthermore, the predicted $P_\mathrm{orb}$ distribution is in broad agreement with the results presented in figure~3 of SC16. It is clear from Figure~\ref{fig:fig1} that the contributions of both early and late cores are essential to the goodness of the fit produced by this model: if only late-emerging cores were considered (as was done in previous treatments of the remnant-core scenario), the distribution would not extend all the way up to $P_\mathrm{orb}\sim2$\,days, as inferred from the data. 

Since the detailed properties of isolated HEs are not yet well determined, we do not attempt to fine-tune the model parameters. Instead, we treat Model~1 as our fiducial case and examine the effect that varying the parameters would have on the model predictions. As expected, the fraction of HJs that cross the RL and are converted into remnant cores increases---resulting in larger ratios of cores (either surviving or destroyed) to surviving HJs and of early to late cores---when the value of $e_\mathrm{0,max}$ goes up or when the eccentricity distribution becomes flatter (corresponding to a lower absolute value of $\beta$): a larger $e_\mathrm{0,max}$ implies a smaller periastron distance for a given value of $a_0$, whereas a lower value of $|\beta|$ results in a larger fraction of planets having a high value of $e_0$. By the same token, when $a_\mathrm{R,p}$ is smaller---corresponding to a lower value of $q$---so that fewer HJs cross the RL, the fraction of surviving cores and that of early cores among them are reduced. However, since the planet's RL is now closer to that of its remnant core and $\tau_\mathrm{d,c}(a)\propto a^{13/2}$ (see Equation~\eqref{eq:a_t}), the fraction of destroyed cores goes up. The behavior of the relative frequencies is similar---lower for surviving cores and higher for destroyed ones---when the value of $Q^\prime_{*,\mathrm{c}}$ is reduced. However, the fraction of late cores among surviving HEs does not increase in this case: in fact, in the example that we present (Models~7 and~8) the tidal dissipation is so efficient that most of the late cores are destroyed and the distribution of surviving HEs is dominated by early remnants. All in all, these results demonstrate that the proposed scenario for the origin of isolated HEs is robust in that it does not depend sensitively on the parameter choices. At the same time, different parameter combinations give rise to distinct distributions, which should in principle make it possible to constrain the model through fits to the data. For example, Models~4 and~6 yield similar ratios of HEs to HJs to that of our fiducial model (see Table~\ref{tab:table1}); however, the three models differ in the details of the predicted $P_\mathrm{orb}$ distribution (see Figure~\ref{fig:fig1}).

\floattable
\begin{deluxetable}{lccccccc}
\tablecaption{Simulation Results}
\tablehead{\colhead{Model No.}
  & \colhead{\% Surviving} & \colhead{\% Surviving}  & \colhead{\% Destroyed}  & \colhead{\% Surviving}  & \colhead{\% Destroyed}  & \colhead{Surviving HEs as}  & \colhead{Destroyed HEs as}\\ \colhead{and $e_0$ Range} & \colhead{HJs} & \colhead{Early HEs} & \colhead{Early HEs} & \colhead{Late HEs} & \colhead{Late HEs}  & \colhead{\% of Surviving HJs} & \colhead{\% of Surviving HJs}}
\startdata
$(1)$\tablenotemark{a},\ $[0.5,0.9]$& 9.4 & 5.1 & 0.0  &  4.0 &  2.3 &  96.2 &  24.0 \\
$(2)$\tablenotemark{a},\ $[0.5,0.95]$&  9.8  &  10.2  &  0.6 &  5.1 &  3.2  &  156.2  & 38.3  \\
\hline 
$(3)$\tablenotemark{b},\ $[0.5,0.9]$& 7.1  &  2.2  &  0.0  &  2.1  &  1.3  &  60.0  &  17.9  \\
$(4)$\tablenotemark{b},\ $[0.5,0.95]$&  7.1  & 4.1  &  0.2  & 2.5  & 1.7  &  93.8  &  26.7  \\
\hline 
$(5)$\tablenotemark{c},\ $[0.5,0.9]$&  10.4  &  2.3  &  0.0  &  3.6  &  4.5 &  57.0  &  43.3  \\
$(6)$\tablenotemark{c},\ $[0.5,0.95]$&  10.8  &  6.1  &  0.5  &  4.6  &  6.8  &  99.7 &  67.7  \\
\hline 
$(7)$\tablenotemark{d},\ $[0.5,0.9]$&  9.5  &  4.6 &  0.5  &  0.4  &  6.0  &  52.5  &  69.4 \\
$(8)$\tablenotemark{d},\ $[0.5,0.95]$&  9.7  &  7.3  &  3.5  &  0.5  &  7.9  &  80.5 &  116.8  \\
\enddata
\tablenotetext{a}{$\beta=0$, $q=3.46$, $Q^\prime_{*,\mathrm{c}}=10^7$.}
\tablenotetext{b}{$\beta=-2$, $q=3.46$, $Q^\prime_{*,\mathrm{c}}=10^7$.}
\tablenotetext{c}{$\beta=0$, $q=2.7$, $Q^\prime_{*,\mathrm{c}}=10^7$.}
\tablenotetext{d}{$\beta=0$, $q=3.46$, $Q^\prime_{*,\mathrm{c}}=10^6(P_\mathrm{orb}/4\,\mathrm{days})$.}
\label{tab:table1}
\end{deluxetable}

\section{Discussion}
\label{sec:discuss}

We have shown that the same process that can explain the occurrence of the sub-Jovian desert in the ($\log{P_\mathrm{orb}},\log{M_\mathrm{p}}$) plane---namely, HEM that brings planets to the vicinity of their respective RLs---can plausibly also account for the population of dynamically isolated HEs that was recently identified in the ($\log{P_\mathrm{orb}},\log{R_\mathrm{p}}$) plane. In this picture, these objects represent the remnant cores of HJs that crossed the RL---either on their original high-eccentricity orbits or following circularization and orbital decay---and were then rapidly stripped of their gaseous envelopes. We found that, for reasonable choices of the model parameters, it is possible to reproduce both the spatial distribution and the occurrence rate inferred for these planets. It is, however, important to stress that these results only serve to place constraints on the underlying process and that an actual physical model is required to demonstrate that HEM can in fact lead to such an outcome in real systems. In a similar vein, our choice of remnant-core sizes was based on the observationally inferred characteristics of HEs and is not grounded in theory. Indeed, on the face of it the identification of remnant cores of HJs with Earth-size planets may seem puzzling in view of the fact that, in the popular core-accretion scenario for the formation of giant planets,  fairly large core masses ($\gtrsim 10\,M_\earth$) are often inferred \citep[e.g.,][]{Rafikov11}. However, dust depletion in the planet formation region could reduce these values  \citep[e.g.,][]{HoriIkoma10,Piso+15}. It is also conceivable that the mass of a rocky remnant is smaller than that of the original core as a result of erosion in liquid hydrogen that could take place under the high-pressure conditions near the center of a Jupiter-size planet \citep[e.g.,][]{Militzer+16}. HEs might thus be useful for constraining both the formation process and the internal structure of giant planets.

In a recent paper, \citet{Winn+17} presented evidence for distinct metallicity distributions for the hosts of USP planets and of HJs, and used this result to argue against the identification of USP planets as the remnant cores of giant planets. These authors derived a formal ($2\sigma$) upper bound of 46\% on the fraction of sampled USP host stars that could have been drawn from the same metallicity distribution as the sampled HJ hosts. This value can be compared with the lower bound on the fraction of members in the USP planet sample that could be identified as isolated HEs, which SC16 estimated to be at least $\sim$17\% and possibly over 40\%. It is noteworthy that even near the higher end of this range the fraction of HEs among USP planets could be lower than the upper bound on the fraction of USP planets with HJ-like hosts, which implies that our identification of the HE progenitors with HJs need not conflict with the finding of \citet{Winn+17}.\footnote{Note that the SC16 estimate of the lower bound on the fraction of members in the \citet{Sanchis-Ojeda+14} sample that could be identified as isolated HEs was based on the corresponding fraction that they inferred for their own sample. However, given that a significant fraction of the HEs identified by SC16 in their sample evidently have $P_\mathrm{orb}>1\,$day and thus lie outside the nominal orbital range of the USP planet sample, the actual fraction of HEs among USP planets could be lower, which would strengthen the just-made consistency argument.} This is not a surprising conclusion given the various mechanisms that could potentially give rise to the observed USP planets, which (as reviewed in \citealt{Winn+17}) include in-situ formation and disk migration of rocky planets as well as the erosion (through photoevaporation or RLO) of the gaseous envelopes of higher-mass planets (either giants or super-Earths). USP planets could possibly comprise several subpopulations, with HEs that are associated with the remnant cores of tidally stripped HJs and have $P_\mathrm{orb}<1$\,day being just one of them. It may thus be expected that when an explicit list of isolated HE candidates becomes available and these planets' properties can be compared with those of the USP candidates studied by \citet{Sanchis-Ojeda+14}, the metallicities of their host stars would turn out to be systematically higher---consistent with the proposed identification of these HEs with remnant cores of HJs---than those of the majority of USP planet hosts. Another implication of this model that could potentially also be checked is that isolated HEs would have a lower probability of having additional companions with $P_\mathrm{orb} < 50$\,days than the overall USP planet population (for which this probability was inferred by \citealt{Sanchis-Ojeda+14} to be significantly higher than for HJs). This probability could possibly also be lower for shorter-period ($P_\mathrm{orb} \la 1$\,day) HEs than for longer-period ones. This is because early-core systems (which predominate among the longer-period HEs) spend most of their lifetimes (typically a few Gyr) without an HJ that could impede the arrival or in-situ formation of short-period companions.

The total number of destroyed cores in our model can be comparable to that of the surviving ones, which raises the question of whether the HJs ingested by the host star might leave an observable imprint on the chemical composition of the stellar photosphere. Using the similarity in the inferred occurrence rates of HEs and HJs, we estimate that $\la$1\% of Sun-like stars would have incorporated refractory material from tidally destroyed HJs into their envelopes through the HEM process that gives rise to HJs and isolated HEs. This value is too small to explain the inference by \citet{Melendez+17}, based on Li abundance measurements in solar twins, that $\sim$15\% of Sun-like stars exhibit ingestion signatures of this type. However, the ``stranded HJ'' scenario \citep{MatsakosKonigl15}, wherein $\sim$50\% of solar-type stars ingest an HJ when they are $\lesssim1$\,Gyr old, could potentially account for this finding.\footnote{The Li depletion process that underlies the observed correlation between Li abundance and age in solar twins could make it difficult to identify anomalous abundance signatures in stars that are observed more than a few billion years after the planet ingestion episode, so some fraction of the systems that harbored a stranded HJ may not exhibit a discernible enhancement in the Li abundance.} In this picture, the source of the observed Li overabundance is the rocky material that comprised the original core of the stranded HJ. This interpretation is consistent with the enhancement in both Li and the refractory elements that was measured in the 6~Gyr~old solar twin HP~68468 by \citet{Melendez+17} and attributed by them to the ingestion of $\sim 6\,M_\earth$ of rocky material when the star was possibly only $\sim 1$\,Gyr old.

\acknowledgements
We thank Jacob Bean, Dan Fabrycky, Benjamin Montet, Leslie Rogers, Jason Steffen, Lauren Weiss, and the referees for helpful discussions or correspondence. This work was supported in part by NASA ATP grant NNX13AH56G.


\begin{thebibliography}{29}
\expandafter\ifx\csname natexlab\endcsname\relax\def\natexlab#1{#1}\fi

\bibitem[{{Clausen} \& {Tilgner}(2015)}]{ClausenTilgner15}
{Clausen}, N., \& {Tilgner}, A. 2015, \aap, 584, A60

\bibitem[{{Essick} \& {Weinberg}(2016)}]{EssickWeinberg16}
{Essick}, R., \& {Weinberg}, N.~N. 2016, \apj, 816, 18

\bibitem[{{Faber} {et~al.}(2005){Faber}, {Rasio}, \& {Willems}}]{Faber+05}
{Faber}, J.~A., {Rasio}, F.~A., \& {Willems}, B. 2005, \icarus, 175, 248

\bibitem[{{Ford} \& {Rasio}(2006)}]{FordRasio06}
{Ford}, E.~B., \& {Rasio}, F.~A. 2006, \apjl, 638, L45

\bibitem[{{Giacalone} {et~al.}(2017){Giacalone}, {Matsakos}, \&
  {K{\"o}nigl}}]{Giacalone+17}
{Giacalone}, S., {Matsakos}, T., \& {K{\"o}nigl}, A. 2017, ArXiv e-prints, arXiv:1708.07543

\bibitem[{{Ginzburg} \& {Sari}(2017)}]{GinzburgSari17}
{Ginzburg}, S., \& {Sari}, R. 2017, MNRAS, 469, 278

\bibitem[{{Guillochon} {et~al.}(2011){Guillochon}, {Ramirez-Ruiz}, \&
  {Lin}}]{Guillochon+11}
{Guillochon}, J., {Ramirez-Ruiz}, E., \& {Lin}, D. 2011, \apj, 732, 74

\bibitem[{{Hori} \& {Ikoma}(2010)}]{HoriIkoma10}
{Hori}, Y., \& {Ikoma}, M. 2010, \apj, 714, 1343

\bibitem[{{Jackson} {et~al.}(2009){Jackson}, {Barnes}, \&
  {Greenberg}}]{Jackson+09}
{Jackson}, B., {Barnes}, R., \& {Greenberg}, R. 2009, \apj, 698, 1357

\bibitem[{{Jackson} {et~al.}(2016){Jackson}, {Jensen}, {Peacock}, {Arras}, \&
  {Penev}}]{Jackson+16}
{Jackson}, B., {Jensen}, E., {Peacock}, S., {Arras}, P., \& {Penev}, K. 2016,
  Celestial Mechanics and Dynamical Astronomy, 126, 227

\bibitem[{{Jia} \& {Spruit}(2017)}]{JiaSpruit17}
{Jia}, S., \& {Spruit}, H.~C. 2017, \mnras, 465, 149

\bibitem[{{Matsakos} \& {K{\"o}nigl}(2015)}]{MatsakosKonigl15}
{Matsakos}, T., \& {K{\"o}nigl}, A. 2015, \apjl, 809, L20

\bibitem[{{Matsakos} \& {K{\"o}nigl}(2016)}]{MatsakosKonigl16}
---. 2016, \apjl, 820, L8

\bibitem[{{Matsumura} {et~al.}(2010){Matsumura}, {Peale}, \&
  {Rasio}}]{Matsumura+10}
{Matsumura}, S., {Peale}, S.~J., \& {Rasio}, F.~A. 2010, \apj, 725, 1995

\bibitem[{{Mel{\'e}ndez} {et~al.}(2017){Mel{\'e}ndez}, {Bedell}, {Bean},
  {Ram{\'{\i}}rez}, {Asplund}, {Dreizler}, {Yan}, {Shi}, {Lind},
  {Ferraz-Mello}, {Galarza}, {dos Santos}, {Spina}, {Maia}, {Alves-Brito},
  {Monroe}, \& {Casagrande}}]{Melendez+17}
{Mel{\'e}ndez}, J., {Bedell}, M., {Bean}, J.~L., {et~al.} 2017, \aap, 597, A34

\bibitem[{{Militzer} {et~al.}(2016){Militzer}, {Soubiran}, {Wahl}, \&
  {Hubbard}}]{Militzer+16}
{Militzer}, B., {Soubiran}, F., {Wahl}, S.~M., \& {Hubbard}, W. 2016, Journal
  of Geophysical Research (Planets), 121, 1552

\bibitem[{{Piso} {et~al.}(2015){Piso}, {Youdin}, \& {Murray-Clay}}]{Piso+15}
{Piso}, A.-M.~A., {Youdin}, A.~N., \& {Murray-Clay}, R.~A. 2015, \apj, 800, 82

\bibitem[{{Rafikov}(2011)}]{Rafikov11}
{Rafikov}, R.~R. 2011, \apj, 727, 86

\bibitem[{{Sanchis-Ojeda} {et~al.}(2014){Sanchis-Ojeda}, {Rappaport}, {Winn},
  {Kotson}, {Levine}, \& {El Mellah}}]{Sanchis-Ojeda+14}
{Sanchis-Ojeda}, R., {Rappaport}, S., {Winn}, J.~N., {et~al.} 2014, \apj, 787,
  47

\bibitem[{{Santerne} {et~al.}(2016){Santerne}, {Moutou}, {Tsantaki}, {Bouchy},
  {H{\'e}brard}, {Adibekyan}, {Almenara}, {Amard}, {Barros}, {Boisse},
  {Bonomo}, {Bruno}, {Courcol}, {Deleuil}, {Demangeon}, {D{\'{\i}}az},
  {Guillot}, {Havel}, {Montagnier}, {Rajpurohit}, {Rey}, \&
  {Santos}}]{Santerne+16}
{Santerne}, A., {Moutou}, C., {Tsantaki}, M., {et~al.} 2016, \aap, 587, A64

\bibitem[{{Seager} {et~al.}(2007){Seager}, {Kuchner}, {Hier-Majumder}, \&
  {Militzer}}]{Seager+07}
{Seager}, S., {Kuchner}, M., {Hier-Majumder}, C.~A., \& {Militzer}, B. 2007,
  \apj, 669, 1279

\bibitem[{{Steffen} \& {Coughlin}(2016)}]{SteffenCoughlin16}
{Steffen}, J.~H., \& {Coughlin}, J.~L. 2016, Proceedings of the National
  Academy of Science, 113, 12023

\bibitem[{{Steffen} \& {Farr}(2013)}]{SteffenFarr13}
{Steffen}, J.~H., \& {Farr}, W.~M. 2013, \apjl, 774, L12

\bibitem[{{Teitler} \& {K{\"o}nigl}(2014)}]{TeitlerKonigl14}
{Teitler}, S., \& {K{\"o}nigl}, A. 2014, \apj, 786, 139

\bibitem[{{Valsecchi} {et~al.}(2015){Valsecchi}, {Rappaport}, {Rasio},
  {Marchant}, \& {Rogers}}]{Valsecchi+15}
{Valsecchi}, F., {Rappaport}, S., {Rasio}, F.~A., {Marchant}, P., \& {Rogers},
  L.~A. 2015, \apj, 813, 101

\bibitem[{{Valsecchi} \& {Rasio}(2014)}]{ValsecchiRasio14}
{Valsecchi}, F., \& {Rasio}, F.~A. 2014, \apjl, 787, L9

\bibitem[{{Valsecchi} {et~al.}(2014){Valsecchi}, {Rasio}, \&
  {Steffen}}]{Valsecchi+14}
{Valsecchi}, F., {Rasio}, F.~A., \& {Steffen}, J.~H. 2014, \apjl, 793, L3

\bibitem[{{Winn} {et~al.}(2017){Winn}, {Sanchis-Ojeda}, {Rogers}, {Petigura},
  {Howard}, {Isaacson}, {Marcy}, {Schlaufman}, {Cargile}, \& {Hebb}}]{Winn+17}
{Winn}, J.~N., {Sanchis-Ojeda}, R., {Rogers}, L., {et~al.} 2017, \aj, 154, 60

\bibitem[{{Wu} \& {Lithwick}(2011)}]{WuLithwick11}
{Wu}, Y., \& {Lithwick}, Y. 2011, \apj, 735, 109

\end{thebibliography}
\end{document}